\pdfoutput=1
\documentclass[aps,prb,floatfix,superscriptaddress,twocolumn,reprint]{revtex4-2}
\usepackage{eurosym}
\usepackage{amsbsy}
\usepackage{latexsym,epsfig,graphicx}
\usepackage{dcolumn}
\usepackage{graphicx}
\usepackage{subfigure}
\usepackage{comment}
\usepackage{color}
\usepackage{bm}
\usepackage{mathrsfs}
\usepackage{amssymb}
\usepackage{amsthm}
\usepackage{amsfonts}
\usepackage{amsmath}
\usepackage{xspace}
\usepackage{epstopdf}
\usepackage{tabularx}
\usepackage{longtable}
\usepackage[colorlinks=true, letterpaper=true, pdfstartview=FitV, linkcolor=blue, citecolor=blue, urlcolor=blue]{hyperref}
\usepackage[normalem]{ulem}
\usepackage[version=4]{mhchem}

\newcommand{\dslash}{\hbox{\kern .22em\raise.18ex\hbox{$/$}\kern-.60em  $\partial$}}

\begin{document}
\title{Dynamical signatures of point-gap Weyl semimetal} 
\author{Haiping Hu}
\email{hhu@iphy.ac.cn}
\affiliation{Beijing National Laboratory for Condensed Matter Physics, Institute of Physics, Chinese Academy of Sciences, Beijing 100190, China}
\affiliation{Department of Physics and Astronomy, George Mason University, Fairfax, Virginia 22030, USA}
\affiliation{Department of Physics and Astronomy, University of Pittsburgh, Pittsburgh, Pennsylvania 15260, USA}
\author{Erhai Zhao}
\email{ezhao2@gmu.edu}
\affiliation{Department of Physics and Astronomy, George Mason University, Fairfax, Virginia 22030, USA}
\author{W. Vincent Liu}
\email{wvliu@pitt.edu}
\affiliation{Department of Physics and Astronomy, University of Pittsburgh, Pittsburgh, Pennsylvania 15260, USA}
\affiliation{Shenzhen Institute for Quantum Science and Engineering, Southern University of Science and Technology, Shenzhen 518055, China}
\affiliation{Shanghai Research Center for Quantum Sciences, Shanghai 201315, China}

\date{\today}

\begin{abstract}
We demonstrate a few unique dynamical properties of point-gap Weyl semimetal, an intrinsic non-Hermitian topological phase in three dimensions. We consider a concrete model where a pair of Weyl points reside on the imaginary axis of the complex energy plane, opening up a point gap characterized by a topological invariant, the three-winding number $W_3$. This gives rise to surface spectra and dynamical responses that differ fundamentally from those in Hermitian Weyl semimetals. First, we predict a time-dependent current flow along the magnetic field in the absence of an electric field, in sharp contrast to the current driven by the chiral anomaly, which requires both electric and magnetic fields. Second, we reveal a novel type of boundary-skin mode in the wire geometry which becomes localized at two corners of the wire cross section. We explain its origin and show its experimental signatures in wave-packet dynamics.
 \end{abstract}
\maketitle

\section{Introduction} Weyl semimetals (WSMs) are three-dimensional (3D) crystals with pairs of isolated band degeneracy points known as the Weyl points (WPs)~\cite{weylreview1,weylreview2,wsmrmp,weylxgwan,weylbalents,weylhuangsm,weyldaixi,weyllv,weylxu}. When the chemical potential lies near the degeneracy points, the low energy quasiparticles are Weyl fermions, i.e., massless chiral fermions obeying the Weyl equation. In the simplest case, a Weyl semimetal has two Weyl points with opposite chirality $\pm 1$ located at $\pm \mathbf{b}$ in momentum space with effective Hamiltonian $H_\pm = \pm v (\mathbf{k}\mp \mathbf{b})\cdot \boldsymbol{s}\pm b_0$. Here $\boldsymbol{s}$ refers to the (pseudo-)spin and $v$ plays the role of the speed of light. The two Weyl points, as the source and drain of Berry flux in momentum space, carry integer topological charge $\pm 1$. This gives rise to a host of fascinating phenomena, including the emergence of gapless excitations in the form of Fermi arcs on surfaces and anomalous Hall effect. Remarkably, WSMs realize the so-called chiral anomaly in quantum field theory~\cite{chiralano1,chiralano2,cmegrushin,cmewsm,cmeyamamoto,cmegoswami,cmefranz}.
For example, in the presence of both $\mathbf{E}$ and $\mathbf{B}$ fields, an effective chiral chemical potential $b_0\propto  \mathbf{E\cdot B}$ is established, leading to an electrical current $\mathbf{j}\propto \mathbf{B}(\mathbf{E\cdot B})$.

Weyl points have been realized and probed in a wide range of physical systems \cite{weyllv,weylxu,weylbook,weylphotonics,weylcold,weylcold2,weylcold3}. In solids, Weyl quasiparticles are often coupled to other degrees of {freedom} such as phonons, magnons, or external fields or bath to acquire finite lifetime \cite{weylfinite1,weylfinite2,weylfinite3}. In recent years, non-Hermitian (NH) Hamiltonians \cite{coll1,coll2,coll3,coll4} have been fruitfully applied to model electronic materials \cite{finite1,finite2,finite3,finite4} and photonic systems with gain and loss \cite{op1,op2,op3,op4,op5,op6,op7,op8}, fueled by the state of the art experimental capability for NH engineering. This motivates us to examine generalized models of WSM as open quantum systems described by NH effective Hamiltonians. The rich, unique topological properties of NH systems can not be captured by the classification framework developed for Hermitian topological band insulators \cite{pclass1,pclass2,pclass3,pclass4,pclass5,pclass6}. Since the energy eigenvalues live on the complex plane, the bands can have point gaps \cite{pointtopo1,pointtopo2,pointtopo3}: the spectrum encloses a simply connected area that contains the reference energy and cannot be smoothly deformed into a gap along the real or imaginary axis. Point gap lies at the heart of a few spectacular properties \cite{coll3,coll4} such as the NH skin effect \cite{ne1,ne2,ne3,ne4,ne5,ne6,ne7,ne8,ne9,ne10,nhsee1,nhsee2,nhsee3,nhsee4,unhse}, where an extensive number of eigenmodes are localized at the boundary.

Recent work has begun to reveal some novel features of NH semimetals \cite{nhsmxu,nhsmzyuzin,nhwsmbudich,nhwsmyang,nhduality,eti,w3invariant,nhfieldtheory,nhsmkawabata,nhsmchinghua,nhsmyangzhesen,nhsmyangzhesen2,nhhowsm,wer2,wer3}. Ref. \cite{nhduality} analyzed a model with 8 {WPs} on the complex energy plane to predict the appearance of skin modes at surfaces perpendicular to an applied magnetic field. Ref. \cite{eti} considered {WPs} with different lifetimes as a limit of exceptional topological insulators and related the emergence of Fermi arcs to a point-gap invariant. Experimentally, a novel kind of Weyl exceptional ring \cite{nhsmxu} has bee realized both in optical waveguides \cite{wer2} and phononic crystals \cite{wer3}. Despite the progress and extensive studies which focus on the static properties of NH topological systems, their dynamical properties remain poorly understood. What are the new and unique effects in dynamics and electromagnetic response dictated by the NH band topology?

\begin{figure}[t!]
\includegraphics[width=3.35in]{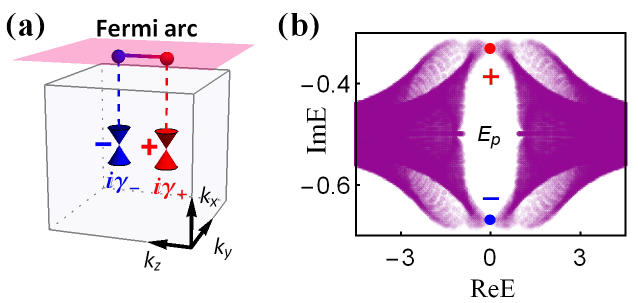}
\caption{Schematics of point-gap WSM. (a) A pair of WPs are split along the $k_z$ axis, leading to surface Fermi arcs. The two WPs carry opposite charges ($\pm$) and have different imaginary energies $i\gamma_{\pm}$, i.e., different dissipation rates. (b) The energy spectra of the lattice model Eq. \eqref{model} on the complex plane. The two WPs are located on the imaginary energy axis. The point gap surrounded by the bulk bands is characterized by invariant $W_3(E_p)=1$ for reference energy $E_p$ inside the point gap. The parameters are $b=0.9$, $\delta=0.2$, $\gamma=-0.5$, $m=3.1$.}\label{fig1}
\end{figure}

In this paper, we investigate a minimal model of NH WSM, with a pair of WPs located on the imaginary axis, $E=i\gamma_{\pm}$, see Fig. \ref{fig1}. The point gap on the complex energy plane dictates the bulk topology and dynamical response. We predict a new effect--time-dependent current induced by magnetic field, $\mathbf{j}(t)\propto \mathbf{B}$, that saturates at long time. This dynamical chiral magnetic effect here differs fundamentally from that in Hermitian {WSM} because it does not require an $\mathbf{E}$ field, is time-dependent and is driven by the different dissipation rates of the WPs. Furthermore, we showcase the existence of a novel type of boundary-skin modes using the Chern number and the spectral winding number, and propose their observation through wave-packet dynamics. 

This paper is organized as follows. In Sec. \ref{secii}, we introduce a minimal model of NH WSM with a pair of EPs of different imaginary energies and demonstrate the existence of point gap and the relevant bulk topological invariants. In Sec. \ref{seciii}, we study the dynamical charge pumping effect in the presence of electromagnetic field. We solve the Landau levels and calculate the pumped charge during time evolution. In Sec. \ref{seciv}, we discuss the boundary-skin modes in wire geometry due to the point-gap topology. In Sec. \ref{secv}, we turn to the wave-packet dynamics as an alternative signature of the point-gap WSM. We conclude in Sec. \ref{secvi} and discuss possible experimental realizations of the point-gap WSM in photonic and condensed matter system. We leave detailed derivations and calculations in the Appendices. Appendix \ref{smA} provides details on our lattice model's spectral windings and symmetries. In Appendix \ref{smB} and Appendix \ref{smC}, we explicitly derive the Landau levels under an orbital magnetic field, and dynamical charge pumping with imaginary Landau levels, respectively. We investigate the surface Fermi arcs as the bulk-edge correspondence of point-gap WSM in Appendix \ref{smD} and the energy spectra and wave-packet dynamics along $z$-wire in Appendix \ref{smE}. In Appendix \ref{smF}, we propose the possible realizations of the lattice Hamiltonian in coupled micro-ring resonators and condensed matter systems. In Appendix \ref{smG}, we discuss the observation of the dynamical effects.

\section{Model Hamiltonians and topological invariants}\label{secii}
Consider a pair of WPs, labeled by subscripts $\pm$ and located at $\mathbf{k}=(0,0,\mp b_z)$ with imaginary energies $E=i\gamma_{\pm}$. They are described by the effective Hamiltonian
\begin{eqnarray}
H_{\pm}=k_xs_x+k_ys_y\pm(k_z\pm b_z)s_z+i\gamma_{\pm}s_0.
\label{lowenergy}
\end{eqnarray}
Here the Pauli matrices $s_j$ with $j=x,y,z$ denote the (pseudo-)spin degrees of freedom and $s_0$ is the identity matrix. The two WPs are separated in momentum space by $2\mathbf{b}=(0,0,{2b_z})$. Note they have opposite chirality $\pm 1$ and different dissipation rates, i.e., inverse lifetimes. For simplicity, we assume the group velocity of the Weyl fermions is isotropic and set $v=1$. We also assume the system overall is dissipative and $\gamma_{\pm}<0$. 

As a concrete example, we consider a four-band lattice model. Its Hamiltonian in momentum space reads
\begin{eqnarray}\label{model}
H_\mathbf{k}=\tau_x\bm a_{\bm k}\cdot\bm \sigma+m_{\bm k}\tau_z\sigma_0+b\tau_0\sigma_z+i \delta \tau_x\sigma_0+i\gamma\tau_0\sigma_0.\;\;\;
\end{eqnarray}
Here the Pauli matrices $\tau_j$  ($\sigma_j$) denote the orbital (spin) degrees of freedom, $\tau_0$ and $\sigma_0$ are identity matrices. The first term with $\bm a_{\bm k}=(\sin k_x,\sin k_y,\sin k_z)$ describes spin-orbit coupling, and $m_{\bm k}=\cos k_x+\cos k_y+\cos k_z-m$. Without the last two NH terms, the model furnishes a prototype of {WSM} \cite{wsmrmp} with a pair of zero-energy WPs separated along the $k_z$ axis. Upon the introduction of $\gamma$  and $\delta$, the two WPs split along the imaginary axis, accompanied by the opening of a point gap inside the bulk bands as depicted in Fig. \ref{fig1}(b). Near the WPs, $H_\mathbf{k}$ reduces to the continuum model Eq. \eqref{lowenergy}, with $b_z$ and $\gamma_{\pm}$ functions of $b$, $m$, and $\delta$, after we rescale the momentum so the group velocity along $x,y,z$ become the same $v$. A more general lattice model was previously introduced in Ref. \cite{eti}. The key features of point-gap WSM  do not depend on the specific lattice model chosen. 

The band topology of $H_\mathbf{k}$ is characterized by a point-gap invariant, the three-winding number
\cite{pclass1,pclass2}
\begin{eqnarray}
W_3(E_p)=-\frac{1}{24\pi^2}\int_{BZ} d^3\bm k~\epsilon^{ijk}\textrm{Tr}[Q_iQ_jQ_k],
\end{eqnarray}
where $E_p$ is a chosen reference energy inside the point gap, $Q_i=(H_\mathbf{k}-E_p)^{-1}\partial_{k_i}(H_\mathbf{k}-E_p)$, and $\epsilon^{ijk}$ is the Levi-Civita symbol. This is possible owing to the existence of a point gap, so that $H_\mathbf{k}$ for each momentum $\bm k$ within the Brillouin zone (BZ)  can be continuously deformed into a unitary matrix \cite{pclass1,pclass2,nhduality,eti,nhfieldtheory}. It can be checked that for our model $W_3(E_p)=1$. To understand the boundary and skin modes in point-gap WSM, two kinds of topological indices of lower dimensions are also needed. Consider a general direction $\hat{l}$, let us label the momentum along $\hat{l}$ as $k_l$ and define transverse momentum $\mathbf{k}_\perp=\mathbf{k}-k_l  \hat{l}$. For fixed values of $\mathbf{k}_\perp$, $H_\mathbf{k}$ defines a 1D Hamiltonian $h_{1D}(k_l)$ where the parametric dependence on $\mathbf{k}_\perp$ is suppressed for brevity. The spectral winding number for $h_{1D}(k_l)$, 
\begin{eqnarray}\label{1dwindingnumber}
w_{l}(E_p)=\frac{1}{2\pi i}\int dk_l\partial_{k_l}[\log\det(h_{1D}(k_l)-E_p)],
\end{eqnarray}
is an integer when $E_p$ lies within the point gap of  $h_{1D}$. In particular, we find $w_{x}=w_{y}=0$, due to the NH time-reversal symmetry \cite{pclass2,cte}: $T_x H(k_x,k_y,k_z)T_x^{-1}=H(-k_x,k_y,k_z)$ and $T_y H(k_x,k_y,k_z)T_y^{-1}=H(k_x,-k_y,k_z)$ where $T_x=\tau_0\sigma_zT$, $T_y=T$ and $T$ stands for transposition. Note the difference from the Hermitian systems, here time-reversal symmetry $T_x$, $T_y$  include the transpose operation. For fixed value of $k_l$, $H_\mathbf{k}$ reduces to a 2D Hamiltonian $h_{2D}(\mathbf{k}_\perp)$. Provided that the bands of $h_{2D}$ at $ReE<0$ and $ReE>0$ are separated, we can define a {\it total Chern number} $C(k_l)$ for all the $ReE<0$ bands. For example, we find $C(k_z)=1$ for  $k_z\in[-b_z,b_z]$ and zero otherwise.

\section{Dynamical charge pumping by magnetic field}\label{seciii}
\begin{figure}[t!]
\includegraphics[width=3.22in]{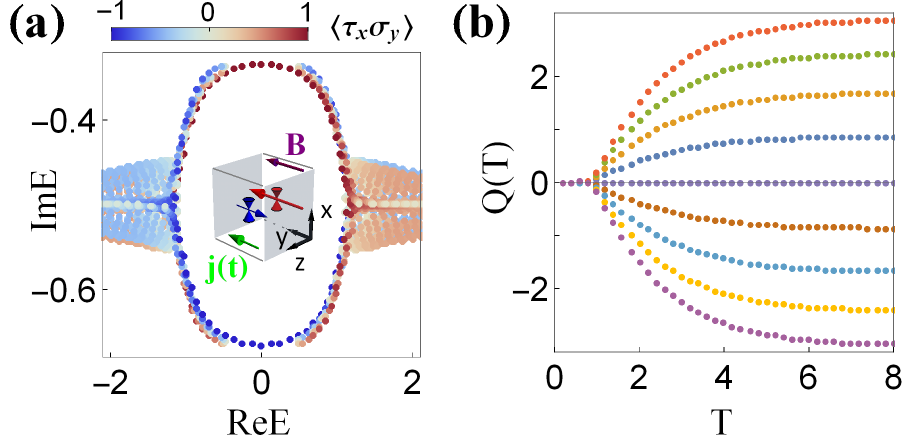}
\caption{Dynamical charge pumping by magnetic field $B$ along $y$. (a) The complex energy spectra for $B={2\pi}/{L_x}$. The color indicates the biorthogonal expectation value $\langle\psi_L|\tau_x\sigma_y|\psi_R\rangle$ for each eigenstate. The two chiral Landau levels in Eq. \eqref{LL0} carry opposite pseudo-spin $s_y=\tau_x\sigma_y$ \cite{SM}. The inset schematic: a net current $j(t)$ arises due to the imbalance of the current carried by the two chiral Landau levels. (b) The total pumped charge $Q(T)$ with respect to time $T$ for $B={2p\pi}/{L_x}$, with $p=-4,-3, ..., 4$ from bottom to top. The parameters are $b=0.9$, $\delta=0.2$, $\gamma=-0.5$, $m=3.1$, and $L_x=50$.}\label{fig2}
\end{figure}

The electromagnetic response of point-gap WSM deviates drastically from Hermitian WSMs. To illustrate this, we first provide an intuitive picture for the chiral magnetic response using the low-energy Hamiltonian Eq. \eqref{lowenergy}. Without loss of generality, suppose the magnetic field is along the $y$ direction with magnitude $B$ \cite{fnote1}. In Landau gauge $\bm A=(0,0,-Bx)$, solving for the eigenvalues of Eq. \eqref{lowenergy} with minimal coupling yields the Landau levels [See detailed derivations in Appendix \ref{smB}]:
\begin{eqnarray}
E^{\pm}_{n=0}&=&\pm k_y + i \gamma_{\pm}; \label{LL0}\\
E^{\pm}_{n\neq 0}&=&\mathrm{sign}(n) \sqrt{k_y^2+2eB|n|}+ i\gamma_{\pm}.\label{LLn}
\end{eqnarray}
Here the superscripts $\pm$ denote the two Weyl nodes, while the subscript $n$ labels the Landau levels. The two zero-th Landau levels $E^{\pm}_0$ are chiral: they have opposite group velocity and different dissipation rate $\gamma_{+}\neq \gamma_{-}$. Thus as time goes on, the difference in dissipation rate sets up a density imbalance of fermions moving in the $y$ and $-y$ direction, resulting {in} a net charge current $j(t)$ along the magnetic field, see the inset of Fig. \ref{fig2}(a). (The $n\neq 0$ levels are particle-hole symmetric and do not contribute to the net current.) More specifically, let us assume at $t=0$, the system is Hermitian ($\gamma_{\pm}=0$), all the Landau levels at negative energies are filled. After the NH terms are turned on, the net current at $t>0$ is [See detailed derivations in Appendix \ref{smC}]
\begin{eqnarray}\label{current}
j(t)=\frac{\Lambda D}{2\pi}\large(e^{2t\gamma_+}-e^{2t\gamma_-}\large),
\end{eqnarray}
where $\Lambda$ is a high-energy cutoff, and $D={BL_xL_z}/{2\pi}$ with $L_{x,z}$ the system length along the $x, z$ direction is the degeneracy of each chiral Landau level. The total charge ``pumped" by magnetic field over time lapse $T$ is 
\begin{eqnarray}
Q_\Lambda(T)=\int_0^T dt j(t)=\frac{\Lambda D}{4\pi}\left[\frac{e^{2\gamma_+T}-1}{\gamma_+}-\frac{e^{2\gamma_-T}-1}{\gamma_-}\right].
\end{eqnarray}
After a long time, it saturates to a finite value 
\begin{eqnarray}
Q_\Lambda(\infty)=\frac{\Lambda D (\gamma_+- \gamma_-)}{4\pi \gamma_+\gamma_-}\propto B |\gamma_+-\gamma_-|
\end{eqnarray}
where in the last step $ |\gamma_+-\gamma_-| \ll  |\gamma_++\gamma_-|$ is assumed. We stress that the current is time-dependent and flows in the absence of electric field. In contrast, in Hermitian {WSM} the current is zero if no electric field is applied \cite{cmewsm}. The accumulation of charge leads to a finite electric polarization $\mathbf{P}\propto \mathbf{B}$ in finite-size samples, which can be taken as a defining signature of point-gap WSM.

More generally, if an electric field of magnitude $\mathcal{E}$ is applied in parallel to $\mathbf{B}$, chiral anomaly also contributes to the current. In this case, the density of left- and right-moving fermions, $N_{\pm}$, can be found to take the form  [See detailed derivations in Appendix \ref{smC}]:
\begin{eqnarray}
N_{\pm}(t)=(\frac{\Lambda D}{2\pi}\pm\frac{e^2\mathcal{E}B}{8\pi^2\gamma_{\pm}})e^{2\gamma_{\pm}t}\mp\frac{e^2\mathcal{E}B}{8\pi^2\gamma_{\pm}}.
\end{eqnarray}
In the limit $\mathcal{E}=0$, it reduces to Eq. (\ref{current}) above by identifying $j(t)= N_+-N_-$ (recall the velocity is set to 1). After a long time, a steady current is achieved,
\begin{eqnarray}
j_\mathcal{E}(t\rightarrow \infty)=-\frac{e^2\mathcal{E}B}{8\pi^2}(\frac{1}{\gamma_+}+\frac{1}{\gamma_-}). 
\end{eqnarray}

Alternatively, we can numerically compute the current induced by magnetic field based on the lattice Hamiltonian Eq. \eqref{model}. Panel \ref{fig2}(a) shows the energy spectra. In the presence of $\mathbf{B}=B\hat{y}$, the original {WPs} are replaced by a pair of highly degenerate chiral modes that fill the Landau gap of size $\sim\sqrt{B}$ to connect the bulk bands with Re$E<0$ and Re$E>0$. Assume the initial state $|\Psi_0\rangle$ is a half-filled trivial insulator with dispersion $\epsilon_{\tau\sigma}(k)=-(\cos k_x+\cos k_y+\cos k_z)$ for each spin and orbital component. The time evolution is governed by the density matrix $\rho(t)=|\Psi(t)\rangle\langle\Psi(t)|$ with the time-evolved state $|\Psi(t)\rangle=e^{-i H t}|\Psi_0\rangle$. The total charge pumped by magnetic field after time lapse $T$ is
\begin{eqnarray}
Q(T)=\frac{1}{L_z}\sum_{k_x,k_z}\int_0^Tdt\int dk_y~\textrm{Tr}[\rho(t)\partial_{k_y}H].
\end{eqnarray}
Here $\partial_{k_y}H$ is the velocity operator along $y$. Fig. \ref{fig2}(b) plots the function $Q(T)$ for different magnetic fields. The saturation value $Q(\infty)$ is proportional to the magnetic field and vanishes for $B=0$, in agreement with the analytical results above. Flipping the magnetic field results in charge pumped to the opposite direction. 

The electromagnetic response of Hermitian WSM can be described by a field theory with action $S=(e/2\pi)^2\int dt d^3\mathbf{r}~\theta(\mathbf{r},t) \mathbf{E\cdot B}$~\cite{cmegrushin,cmewsm,cmeyamamoto,cmegoswami,cmefranz}. Here the axion field $\theta(\mathbf{r},t)=2(\mathbf{b\cdot r}-b_0t)$ is linear in the separation of WPs in energy and momentum $b^\mu=(b_0, \mathbf{b})$ with natural units $c=\hbar=1$. It predicts the chiral magnetic effect, i.e., a current $\mathbf{j}=-(e^2/2\pi^2)b_0\mathbf{B}$ which vanishes in equilibrium with $b_0=0$. Attempt to generalize the field theory to point-gap WSM is hampered by an obstacle: the divergence of the Fujikawa integral even for small NH perturbations such as $\gamma_{\pm}$. Thus the dynamical chiral magnetic response found here cannot be explained by analytically continuing $\mathbf{j}=-(e^2/2\pi^2)b_0\mathbf{B}$ via $2b_0=i(\gamma_+-\gamma_-)$. The failure of this formula illustrates that we are dealing with a genuinely novel effect \cite{fnote2}. The theory developed in Ref. \cite{nhfieldtheory} cannot be applied here either, because the charge U(1) symmetry assumed in Ref. \cite{nhfieldtheory} is broken by the NH terms in Eq. \eqref{lowenergy}. 

\begin{figure}[t!]
\includegraphics[width=3.22in]{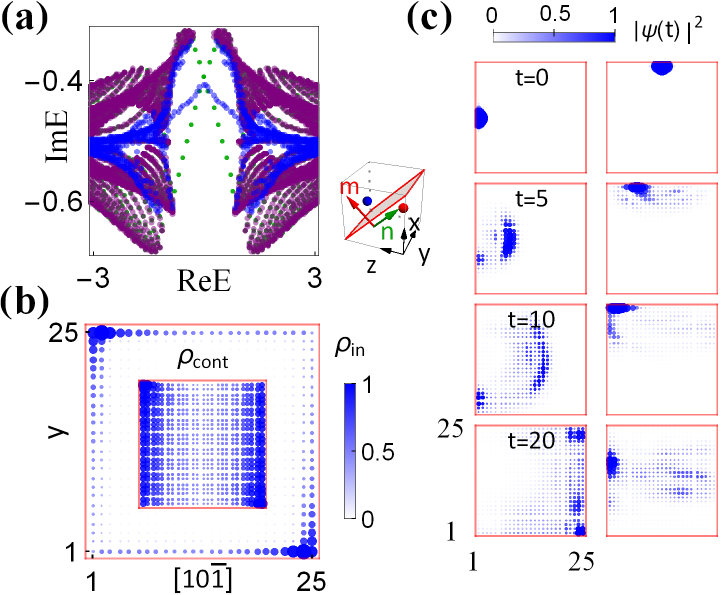}
\caption{Boundary-skin modes and wave-packet dynamics for a wire along $\hat{m}=[101]$. (a) Energy spectra with open (in blue) versus periodic (in purple) boundary conditions in the $\hat{y}$ and $\hat{n}=[10\bar{1}]$ direction, showing the emergence of in-gap modes for open boundaries. The in-gap modes in green are obtained under open $y$ and periodic $\hat{n}$ boundary. The inset illustrates the wire with cross-section boundaries (red lines). (b) Total probability distribution $\rho_{in}$ of the corner-localized in-gap modes. (Middle inset) Total probability distribution of all other modes, which exhibit skin effect along $\hat{n}$. (c) The time evolution of a wave-packet initially localized at site $(1,13)$ (left panel) and $(13,25)$ (right panel) of the cross section that measures $25\times 25$. The wave-packet has width $W_1^2=1$, $W_2^2=6$. The spinor wave function is $|\xi_0\rangle=(1,1,0,0)^T$. The lattice momentum along the wire is chosen as $k_{m}=-0.25$. Other parameters are the same as Fig. \ref{fig1}.}\label{fig3}
\end{figure}

\section{Boundary-skin modes in wire geometry}\label{seciv}
The nontrivial bulk topology leads to the appearance of Fermi-arc surface states that fill the entire point gap \cite{eti}. In Appendix \ref{smD}, we studied the in-gap Fermi arcs for different surface terminations. It also manifests in the emergence of a novel type of boundary-skin modes when the semimetal is cleaved to have intersecting surface planes. Consider for example a wire with a rectangular cross section and extending in the $[101]$ direction $\hat{m}=\hat{x}+\hat{z}$ (red arrow, insets of Fig. \ref{fig3}). For convenience, we label the $[10\bar{1}]$ directions as $\hat{n}=\hat{x}-\hat{z}$, so $(\hat{y},\hat{m}, \hat{n})$ are orthogonal to each other. The spectra of the wire for different boundary conditions are compared in Fig. \ref{fig3}(a) for a particular value of $k_m=\mathbf{k}\cdot\hat{m}$. Shown in color purple is the spectrum for periodic boundary conditions along $\hat{y}$ and $\hat{n}$, and color blue is for open $\hat{y}$ and $\hat{n}$ boundaries where the in-gap modes are visible. It turns out that these in-gap modes are concentrated around two corners of the cross section, according to their total probability distribution $\rho_{in}(i,j)=\sum_{q}|\psi_q(i,j)|^2$ shown in Fig. \ref{fig3}(b). Here $(i,j)$ labels the sites, $q$ labels the in-gap modes, and $\rho_{in}$ is rescaled to have maximum $1$. As $k_m$ is varied, the spatial distribution of these corner modes evolves smoothly, e.g. it is extended for $k_m=0$ and localizes at two other corners as $k_m$ switches sign. Clearly, they are distinct from the chiral edge modes in Chern insulators and cannot be described by the Chern number $C(k_m)$ [defined below Eq. (4)] alone. For open boundaries, the continuum modes with energies overlapping with the bulk spectrum are pushed to localized at the left and right edge, as shown by their total probability distribution $\rho_{cont}(i,j)$ in the middle inset of Fig. \ref{fig3}(b). An extensive number of continuum modes residing near the boundary is known as the NH skin effect. Here the skin effect depends on the orientation/geometry of the surfaces. For example, the skin effect is absent for a $z$-wire with open $x,y$ boundaries [See Appendix \ref{smE} for details]. This is due to the vanishing of the 1D spectral winding $w_{x}=w_y=0$ protected by the NH symmetries $T_x$ and $T_y$. For a given $k_z$, the 2D Hamiltonian $H_{2D}(k_x,k_y)$ describes a non-Hermitian Chern insulator, with the chiral edge modes revealed from the Chern number $C(k_z)$.

We now show that these ``corner modes" can be understood as chiral edge states under the spell of 1D skin effect. Let us start from a point-gap WSM with two open surfaces at $y=1,L$ and periodic in the two other directions $\hat{m}$ and $\hat{n}$. This realizes a 2D slab described by Hamiltonian $h_{2D}(k_{m}, k_{n})$. Its spectrum, shown in green in Fig. \ref{fig3}(a) for a given $k_m$, features two chiral edge modes at $y=1,L$ respectively that cross the bulk gap and disperse with $k_{n}$. Note that for given $k_{m}$, $h_{2D}$ can be regarded as a 1D effective Hamiltonian $h_{1D}(k_n)$. $h_{1D}$ has point gaps on the complex energy plane, and the corresponding 1D spectral winding number $w_{n}$ along the direction $\hat{n}$ is finite, giving rise to 1D skin effect. Thus, upon opening up two additional boundaries normal to $\hat{m}$, the skin effect leads to further localization of the surface modes to the left/right corner. These ``corner modes" [in blue, Fig. \ref{fig3}(a)] indeed reside within the point gap of $h_{1D}(k_n)$. We call them ``boundary-skin modes" because they derive from the chiral edge modes of NH Chern insulators due to the 1D skin effect. Since the finite Chern number is in turn derived from $W_3$, the emergence of boundary-skin modes observed in Fig. \ref{fig3}(a) and (b) can serve as signatures of point-gap WSM. We note the number of boundary-skin modes, bulk skin modes and chiral edges state scale with system size as $L$, $L^2$, and $L$, respectively. 

\section{Wave-packet dynamics}\label{secv}
Besides dynamical charge pumping, we propose an alternative route to extract the topological signatures of point-gap WSM from wave-packet dynamics which can be performed in photonics experiments \cite{nhsee3}. Let $(x_1, x_2)$ be the coordinates within the cross-section area in the wire geometry. At time $t=0$, we prepare a Gaussian wave packet localized at $(a_1,a_2)$ of zero velocity in the plane
\begin{eqnarray}
|\psi_0\rangle=N_0e^{-{(x_1-a_1)^2}/{W_1^2}-{(x_2-a_2)^2}/{W_2^2}}e^{i k_lx_l}|\xi_0\rangle,
\end{eqnarray}
where $W_{1,2}$ are the width of the packet, $N_{0}$ is the normalization factor, $|\xi_0\rangle$ denotes the spinor part of the wave function, and $k_l$ is the momentum along the wire at a fixed value. Fig. \ref{fig3}(c) depicts the time evolution of a wave packet in the cross section of a $[101]$-wire. The left panel shows that the wave packet initially residing near the middle point of the $y$-edge travels directly through the bulk to reach the opposite edge. This occurs because the wave packet has large overlap with the skin modes that reside on the $y$-edge [see Fig. \ref{fig3}(b)], but negligible overlap with the in-gap states which are more concentrated around the corners. The skin modes are not completely localized, giving the wave packet the chance to permeate into the bulk. While for a wave packet initially on the $[10\bar{1}]$-edge (right panel), it first moves counter-clockwise along the edges and starts to permeate into the bulk more significantly once it arrives at the $y$-edge. The evolution dynamics is distinct from that of a $z$-wire, where the wave-packet moves chirally along the edges of the cross section and does not go into the bulk, see numerical simulations in Appendix \ref{smE}]. Thus, the existence of boundary-skin modes can be inferred from the wave-packet dynamics.

\section{Conclusion and discussion}\label{secvi}
To conclude, we predict dynamical charge pumping and boundary skin modes as unique features of NH WSM and attribute them to the point-gap topology and non-Hermicity. These phenomena have no analogs in Hermitian semimetals and cannot be described by the previous field theory framework. Our work lays a foundation for future experiments to explore the dynamics of NH semimetals. The dynamical effects do not rely on fine-tuning to a specific energy window and are more feasible to identify for simulations in photonic and cold atomic platforms. It is straightforward to extend the analysis to other types of topological semimetals \cite{wsmrmp,tdp}. For example, by setting $\bm a_{\bm k}=(\sin k_x\sin k_y,\cos k_y-\cos k_x,\sin k_z)$, we obtain a double-charged NH WSM with point-gap invariant $W_3(E_p)=2$. The lattice Hamiltonian can, in principle, be implemented in photonic lattices and electrical metamaterials \cite{zhao,RLCmeasure1,RLCmeasure2}. As detailed in Appendix \ref{smF} and \ref{smG}, we propose a realization of the lattice Hamiltonian Eq. \eqref{model} using micro-ring resonator arrays with losses, where the couplings (both phase and amplitude) between neighboring resonators can be controlled independently through intermediate waveguides \cite{ringcavity1,ringcavity2,ringcavity3,ringcavity4}. In condensed matter systems, the non-Hermitian dissipation terms can be implemented either through a tailored orbital-dependent coupling with a lossy mode or electron-phonon scattering \cite{eti}.

\begin{acknowledgments}
This work is supported by AFOSR Grant No. FA9550-16-1-0006 (HH, EZ and WVL), NSF Grant No. PHY-2011386 (HH and EZ), the start-up grant of IOP-CAS (H.H.), and the MURI-ARO Grant No. W911NF17-1-0323 through UC Santa Barbara and the Shanghai Municipal Science and Technology Major Project (Grant No. 2019SHZDZX01) (WVL).
\end{acknowledgments}

\appendix

\section{Spectral windings and non-Hermitian symmetry}\label{smA}
The lattice Hamiltonian $H_\mathbf{k}$ (see model \eqref{model} in the main text) contains both Hermitian and non-Hermitian terms. The Hermitian part describes a prototype Weyl semimetal (WSM) with a pair of Weyl points (WPs) inside the $k_z$ axis. The non-Hermitian terms further splits the two WPs along the imaginary axis. Such WP configuration breaks time-reversal symmetry; however if we consider the one-dimensional (1D) Hamiltonian $h_{1D}(k_x)$ with fixed $(k_y,k_z)$ momentum or $h_{1D}(k_y)$ with fixed $(k_y,k_z)$ momentum, the lattice Hamiltonian respects the following non-Hermitian time-reversal symmetry \cite{pclass2,cte}
\begin{eqnarray}
T_x H(k_x,k_y,k_z)T_x^{-1}&=&H(-k_x,k_y,k_z),\\
T_y H(k_x,k_y,k_z)T_y^{-1}&=&H(k_x,-k_y,k_z),
\end{eqnarray}
where $T_x=\tau_0\sigma_zT$, $T_y=T$ and $T$ represents for transposition. The symmetry $T_x$ (or $T_y$) relates the $(100)/(\bar{1}00)$ (or $(010)/(0\bar{1}0)$) surfaces to each other and rules out the skin effect along $x$ (or $y$) direction. To visualize this, we plot the energy spectra along each momentum direction, while keep the other two momenta fixed. As depicted below in Fig. \ref{figs1}(a)(b), the spectra by varying $k_x$ or $k_y$ trace open arcs on the complex plane, indicating the absence of skin modes once open boundary along $x$ or $y$ direction is taken. The spectra by varying $k_z$ form closed loops. For an open $z$-boundary, the extended modes under periodic boundary condition would collapse into skin mode \cite{pointtopo1,pointtopo2,pointtopo3}. Further, the presence or absence of skin modes under open boundary can be verified from the 1D winding number along the corresponding momentum direction. Due to the above non-Hermitian time-reversal symmetry, $W_{1x}(E_p)=W_{1y}(E_p)=0$. While $W_{1z}(E_p)$ can be nonzero when the reference energy $E_p$ is suitably chosen [see Fig. \ref{figs1}(c)].
\begin{figure}[h!]
\includegraphics[width=3.3in]{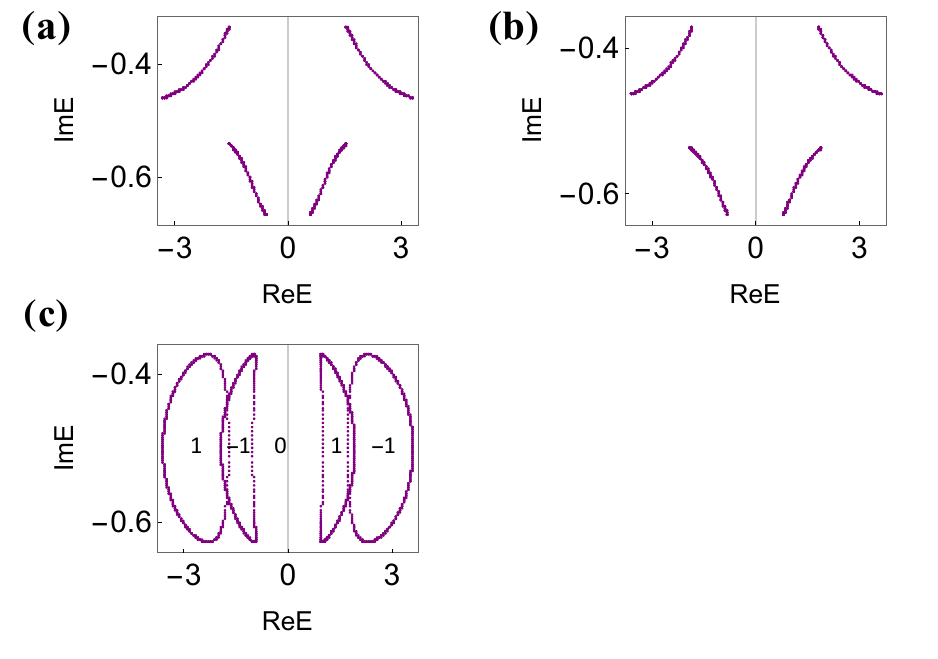}
\caption{Energy spectra of the lattice Hamiltonian $H_\mathbf{k}$ (model (\ref{model}) in the main text) on the complex plane. (a) Open-arc spectra by varying $k_x$ with fixed $k_y=0.5$, $k_z=0.5$. (b) Open-arc spectra by varying $k_y$ with fixed $k_x=0.9$, $k_z=0.5$. (c) Closed-loop spectra by varying $k_z$ with fixed $k_x=0.9$, $k_y=0.5$. The 1D winding number $W_{1z}(E_p)$ is labeled when the reference energy $E_p$ is chosen inside the corresponding spectral region.}\label{figs1}
\end{figure}

\section{Chiral Landau levels with an applied magnetic field}\label{smB}
In the presence of a background magnetic field (For neutral atoms, the magnetic field can be mimicked utilizing the synthetic gauge field technique), the Weyl Hamiltonian coupled to a gauge field is obtained through replacing $\bm k\rightarrow \bm k-e\bm A$. For a magnetic field along $y$ direction, we take the gauge potential $\bm A=(0,0,-B x)$. The low-energy Hamiltonian near the two WPs with opposite charge $\pm1$ (or chirality) reads ($\hbar=c=1$)
\begin{eqnarray}\label{lowenergyB}
H_{\pm}(B)=k_xs_x+k_ys_y\pm (k_z+e Bx\pm b_z)s_z+i\gamma_{\pm}s_0.\notag\\
\end{eqnarray}
We take the $+1$ Weyl node with imaginary energy $\gamma_+$ as an example. Squaring the Hamiltonian yields
\begin{eqnarray} 
[H_{+}(B)-i\gamma_+s_0]^2=k_y^2+k_x^2+(k_z+e B x+b_z)^2-eBs_y.\notag\\
\end{eqnarray}
Note the motion in the $xz$ plane (perpendicular to $\bm B$) is exactly described by the quantum harmonic oscillator, except with the minimum of the potential shifted in coordinate space. The Landau quantization in the $xz$-plane leads to the familiar levels
\begin{eqnarray}
(E_+-i\gamma_+)^2=eB(2n+1)+k_y^2-eBs_y~(n=0,1,2,...),\notag\\
\end{eqnarray}
each with degeneracy $D=\frac{eBL_xL_z}{2\pi}$. The last term (Zeemann splitting) depends on the spin polarization along the magnetic-field direction. When $s_y=+1$ and $n=0$, we get the zero-th Landau level in the main text with linear dispersion
\begin{eqnarray}
E_{0+}=k_y+i\gamma_+.
\end{eqnarray}
While for $n\geq 1$, The $n$-th states of $s_y=-1$ are degenerate with the $(n+1)$-th states of $s_y=+1$. They together constitute the higher Landau levels in the main text, with dispersion
\begin{eqnarray}
E_{n+}=\pm\sqrt{k_y^2+2eBn}+i\gamma_+~~(n\geq1).
\end{eqnarray}
It is worth to mention, only the zero-th Landau level has definite spin polarization along the magnetic field; while the higher Landau levels are constituted of both polarization components, with degeneracy $2D$. Similarly, for the $-1$ Weyl node with imaginary energy $\gamma_-$, the zero-th Landau level has spin polarization $s_y=-1$ and dispersion
 \begin{eqnarray}
 E_{0-}=-k_y+i\gamma_-.
 \end{eqnarray}

 \section{Dynamical charge pumping with imaginary Landau levels}\label{smC}
\begin{figure}[b!]
\includegraphics[width=1.8in]{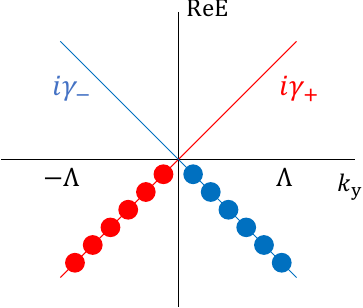}
\caption{Schematics of the zero-th Landau levels with linear dispersions along $k_y$. The right-moving (red) and left-moving (blue) fermions have imaginary energy $i\gamma_+$ and $i\gamma_-$, respectively. $\Lambda$ is the momentum cutoff. At time $t=0$, all the Re$E$ levels are filled (solid dots). }\label{figsm}
\end{figure}
We start from the zero-th Landau levels, which are chiral and possess different dissipation rates as depicted in Fig. \ref{figsm}. The chiral Landau levels emerged under a magnetic field produce a time-dependent parallel current. To see this, we calculate the amount of charge pumped over time lapse $T$. We suppose the system at $t=0$ fill all the Landau levels (i.e., Dirac sea) of Re$E<0$ and denote the initial state as $|\Psi_0\rangle$. The subsequent time-evolution $|\Psi(t)\rangle=e^{-i Ht}|\Psi_0\rangle$ is non-unitary and governed by the density matrix
\begin{eqnarray}
\rho(t)&=&|\Psi(t)\rangle\langle\Psi(t)|\notag\\
&=&\sum_{m,n}e^{-i(E_n-E^*_m)t}|\phi_n\rangle\langle\phi_n|\Psi_0\rangle\langle\Psi_0|\phi_m\rangle\langle\phi_m|,
\end{eqnarray}
where $|\phi_n\rangle$ denotes the eigenfunction of the corresponding Landau level. We set the momentum cutoff as $\Lambda$. The time-dependent current along the magnetic field is then
\begin{eqnarray}
j(t)=\int_{-\Lambda}^{\Lambda}\frac{dk_y}{2\pi}~\textrm{Tr}[\rho(t)\partial_{k_y}H].
\end{eqnarray}
Here $\partial_{k_y}H$ is the particle velocity along the magnetic field. The $\frac{1}{2\pi}$ factor is the density of state. As the higher Landau levels are symmetric with respect to the $k_y$ axis, only the chiral Landau levels contribute to the current. The time-dependent current is simply given by
\begin{eqnarray}
j(t)=\frac{\Lambda D}{2\pi}\large[e^{2\gamma_+t}-e^{2\gamma_-t}\large].
\end{eqnarray}
We can clearly see $j(t)$ is the net current coming from both the left- and right-movers. The total pumped charge during time $T$ is 
\begin{eqnarray}
Q_{\Lambda}(T)=\int_0^T dt~j(t)=\frac{\Lambda D}{4\pi}\large[\frac{e^{2\gamma_+T}-1}{\gamma_+}-\frac{e^{2\gamma_-T}-1}{\gamma_-}\large].\notag\\
\end{eqnarray}

In the following, we provide a field-theory perspective of the dynamical current. The dynamical charge pumping is due to interplay of non-Hermiticity and the chiral Landau levels. We restrict to the zero-th Landau levels with opposite chirality and denote the corresponding field operator describing the chiral fermions as $\chi(t,y)$. In this notation, we have incorporate the $(x,z)$-dependence into $\chi(t,y)$. The effective (1+1)D action describing the two chiral landau levels is
\begin{eqnarray}\label{action}
S=\int dt dy~ i\bar{\chi}(t,y)[\dslash-\frac{\gamma_+-\gamma_-}{2}\gamma^1-\frac{\gamma_++\gamma_-}{2}\gamma^0]\chi(t,y).\notag\\
\end{eqnarray}
Here we have utilized the notation of gamma matrices as $\gamma^0=\sigma_x$, $\gamma^1=-i\sigma_y$ and $\gamma^5=\gamma^0\gamma^1=\sigma_z$, which obey the Clifford algebra $\{\gamma^{\mu},\gamma^{\nu}\}=2g^{\mu\nu}$ in signature $(1,-1)$. 

The field $\chi(t,y)$ can be decomposed into two chiral components $\chi_{\pm}(t,y)=\frac{1}{2}(1\pm\gamma^5)\chi(t,y)$, corresponding to different eigenvalues of $\gamma^5$. In terms of $\chi_{\pm}(t,y)$, the action reads
\begin{widetext}
\begin{eqnarray}
S=\int dt dy~ i[\chi_+^{\dag}(t,y)(\partial_t+\partial_y-\gamma_+)\chi_+(t,y)+\chi_-^{\dag}(t,y)(\partial_t-\partial_y-\gamma_-)\chi_-(t,y)].
\end{eqnarray}
\end{widetext}
Without the dissipation terms, the action (\ref{action}) has both the charge and chiral U(1) symmetry, indicating the conservation of gauge current $j^\mu=\bar{\chi}\gamma^{\mu}\chi$ and chiral current $j^{\mu}_5=\bar{\chi}\gamma^{\mu}\gamma^5\chi$ in classical level. In terms of the two chiral components, $j^0=\chi^{\dag}_+\chi_++\chi^{\dag}_-\chi_-\equiv N_++N_-$ measures the total density of right- and left-moving fermions; and $j^1=\chi^{\dag}_+\chi_+-\chi^{\dag}_-\chi_-\equiv N_+-N_-$ measures their density difference (or current). Vice versa for $j^{\mu}_5$, $j_5^0= N_+-N_-$ and $j_5^0= N_++N_-$ respectively measures their density difference and total density. The existence of the dissipation terms breaks both symmetries, leading to the non-conservation for both the left- and right-movers.

The equation of motion extracted from action (\ref{action}) is
\begin{eqnarray}
\dslash\chi-\large[\frac{\gamma_++\gamma_-}{2}\gamma^0+\frac{\gamma_+-\gamma_-}{2}\gamma^1\large]\chi=0.
\end{eqnarray}
The solutions are given by
\begin{eqnarray}
\chi_+(t,y)=(t-y)e^{\gamma_+t};~~~\chi_-(t,y)=(t+y)e^{\gamma_- t}.
\end{eqnarray}
We can clearly see their physical meaning: $\chi_{\pm}$ represents for the right/left-moving fermions with damping rate $\gamma_{\pm}$, respectively. The fermion density operator satisfies the following damping relation:
\begin{eqnarray}
\partial_t N_+=2\gamma_+N_+;~~~\partial_t N_-=2\gamma_-N_-.
\end{eqnarray}
The fermion density of the right- and left-movers are then $N_+(t)\propto e^{2\gamma_+ t}$ and $N_-(t)\propto e^{2\gamma_- t}$. As the two chiral components move in opposite directions ($y$ and $-y$), their density difference $j^1(t)=j_5^0(t)=N_+(t)-N_-(t)$ induces a net current proportional to $(e^{2\gamma_+t}-e^{2\gamma_-t})$ along the magnetic field, which coincides with the previous density-matrix calculations.

It is worth to mention the case when an additional electric field $\mathcal{E}$ parallel to the magnetic field $\bm B$ is applied. As is well known in quantum field theory, the electric field would induce the chiral anomaly, which breaks the chiral symmetry in the quantum level. The chiral anomaly shifts the density of right- and left-movers by $\pm\frac{e\mathcal{E}}{2\pi}$, respectively. Taking into account this effect, we arrive at the following relation:
\begin{eqnarray}
\partial_t N_{\pm}=\pm\frac{e^2\mathcal{E}B}{4\pi^2}+2\gamma_{\pm}N_{\pm}.
\end{eqnarray}
The solutions are given by
\begin{eqnarray}
N_{\pm}(t)=(N_{0\pm}\pm\frac{e^2\mathcal{E}B}{8\pi^2\gamma_{\pm}})e^{2\gamma_{\pm}t}\mp\frac{e^2\mathcal{E}B}{8\pi^2\gamma_{\pm}}.
\end{eqnarray}
Here $N_{0,\pm}$ is the initial fermion density for the right (+) and left (-) movers, respectively. For the initial configuration depicted in Fig. \ref{figsm} with momentum cutoff $\Lambda$, $N_{0\pm}=\frac{\Lambda D}{2\pi}$. It is easy to see:\\
\noindent \underline{Case (i):} When $\gamma_+=\gamma_-=0$, i.e., no dissipation for both the left- and right-movers, $\partial_t j^0=0$, $\partial_tj^1=\frac{e^2\mathcal{E}B}{2\pi^2}$, which returns to the well-known chiral anomaly. The total particle density is conserved, however, the chiral density is not conserved.\\
\noindent \underline{Case (ii):} When $\gamma_+=\gamma_-\neq0$, i.e., the left- and right-movers have the same dissipation rate, $j^1(t)=\frac{e^2\mathcal{E}B}{4\pi^2\gamma_+}(e^{2\gamma_+t}-1)$. When the electric field $\mathcal{E}=0$, the net current is zero.\\
\underline{Case (iii):} When $\gamma_+\neq\gamma_-\neq0$ and $\mathcal{E}=0$, i.e., without the electric field, $j^1(t)=\frac{\Lambda D}{2\pi}(e^{2\gamma_+t}-e^{2\gamma_-t})$, which is consistent with the previous density-matrix discussions. Even without electric field, a time-dependent current is induced due to the dynamical imbalance between left- and right-movers.\\
\underline{Case (iv):} When $t$ is very large, the competition between the non-Hermitian dissipation and electric-field driving is balanced. And we arrive at the steady-state solution: $N_{\pm}(t\rightarrow\infty)=\mp\frac{e^2\mathcal{E}B}{8\pi^2\gamma_{\pm}}$.

\section{Anisotropic surface Fermi arcs}\label{smD}
\begin{figure}[h!]
\includegraphics[width=3.3in]{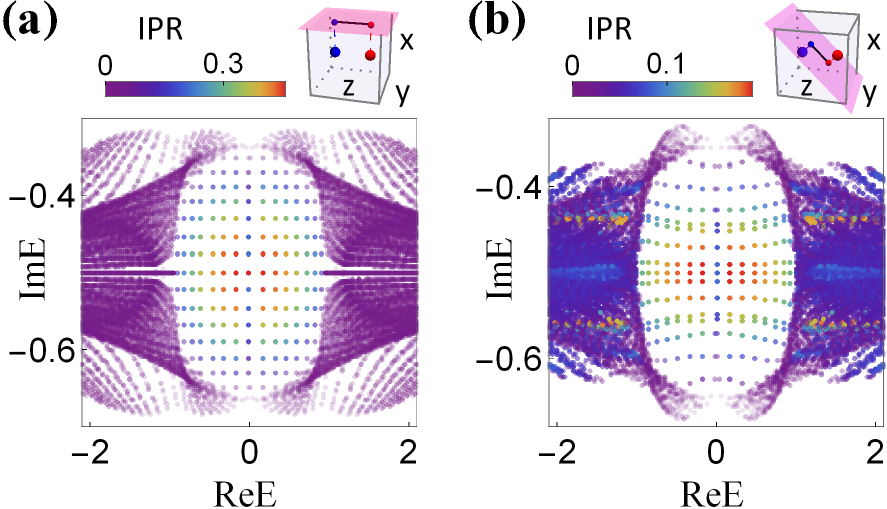}
\caption{Complex Fermi arcs and skin effect at open surfaces. (a) Energy spectra for a point-gap WSM with two open surfaces (see  the pink surface in inset) normal to the $x$ direction, separated by distance $L_x=25$. The colors indicate the Inverse Participation Ratio (IPR) that measures the localization of eigenstates. The in-gap modes consist of Fermi arcs to fill the entire point gap. (b) Same as (a) but for open surfaces normal to the $[10\bar{1}]$ direction with $L_{[10\bar{1}]}=25$. The IPR shows skin effect, i.e., an extensive number of continuum modes (outside the point gap) become localized near the surfaces. The parameters are $b=0.9$, $\delta=0.2$, $\gamma=-0.5$, $m=3.1$.}\label{figs3}
\end{figure}

The bulk-boundary correspondence in point-gap WSM is more complicated than the Hermitian case. This is partly due to the appearance of skin modes which depends on the orientation of the surfaces. We first focus on one of the key signatures of WSM, Fermi arcs on open surfaces.  Fig. \ref{figs3}(a) shows the spectra for open boundaries at {$x=1,L$} (the pink surface parallel to the Weyl node separation in the inset) obtained from numerical solution of the lattice model. Owing to the point-gap invariant {$W_3(E_p)=1$}, surface modes emerge inside the point gap. Here, for clarity, only the spectra of a few dozen discrete values of transverse momenta $\mathbf{k}_\parallel=(k_y,k_z)$ are shown. The surface modes become close-packed to fill the entire point gap region if all $\mathbf{k}_\parallel$ are included. Consider for example the zero-energy surface states at $k_{y,z}=0$, whose wave functions can be found analytically. (The solution is provided at the end of this section.) The complex energy spectrum for small values of $k_{y,z}$ is given by
\begin{eqnarray}
E_s(\mathbf{k}_\parallel) \propto \pm k_y + i\alpha k_z,\label{surfacestate}
\end{eqnarray}
where $\pm$ is for the surface at {$x=1$} and $L$ respectively and $\alpha$ depends on system parameters.  At zero chemical potential, $\mu=$Re$E_s=0$, the surface modes disperses as $E_s\propto ik_z$ and form a continuum with varying Im$E_s$ for {$k_z\in [-b_z,b_z]$} to connect the two {WPs}, i.e., a complex Fermi arc. Fermi arcs for other values of $\mu$ are obtained similarly by solving Re$E_s(\mathbf{k}_\parallel)=\mu$. The in-gap modes can be viewed as a collection of Fermi arcs. Remarkably, together they form a single-sheet ``handkerchief" on the complex $E$ plane, covering the ``hole" of the point gap area exactly once (recall $W_3=1$ in our model). More generally, one can prove that the complex Fermi arcs cover the point-gap area {$W_3(E_p)$} times \cite{eti}. 

In comparison, Fig. \ref{figs3}(b) depicts the spectra for the {$(10\bar{1})/(\bar{1}01)$} open surfaces perpendicular to the diagonal {$\hat{x}-\hat{z}$.} The false color represents the inverse participation ratio (IPR) that measures the wave function localization
\begin{eqnarray}
\textrm{IPR}[|\psi\rangle]=\sum_{j}|\langle j|\psi\rangle|^4.
\end{eqnarray}
Here $j$ labels the lattice layers along $\hat{x}-\hat{z}$, and a high IPR value indicates the localization of wave function $|\psi\rangle$ near the two open surfaces. While the complex Fermi arcs fill the point gap, certain states with energies belonging to the continuum bulk bands have appreciable IPR, i.e., they are pushed from the bulk to localize near the surfaces. This is an example of  non-Hermitian skin effect and it can be understood by analyzing $h_{1D}(k_l)$ with $\hat{l}=\hat{x}-\hat{z}$. Skin effect occurs whenever the spectral windings along $\hat{l}$, as defined in Eq. \eqref{1dwindingnumber}, is nonzero. We can check that $w_{l}(E_p)$ is indeed finite for certain $E_p$ outside the point gap region, in agreement with Fig. \ref{figs3}(b). Note the skin effect depends on the orientation of the open surface. For the $x$-open boundary shown in Fig. \ref{figs3}(a), all the continuum states remain extended. Skin effect is absent in this case because spectral windings along the $x$ and $y$ direction vanish, $w_{x}=w_{y}=0$. \\

\noindent\textit{Solution of the surface states Eq. (\ref{surfacestate})}:\\
For the lattice Hamiltonian $H_\mathbf{k}$ (see Eq. (\ref{model}) in the main text), when the $x$ direction is open, $k_y$ and $k_z$ are good quantum numbers and surface states emerge inside the point gap. We first consider the special case with $k_y=k_z=0$. The surface states can be either on the $(100)$ or $(\bar{1}00)$ surfaces. To proceed, we rewrite the tight-binding form of Hamiltonian $H_\mathbf{k}$ along $x$ direction (the constant non-Hermitian term $i\gamma\tau_0\sigma_0$ is dropped off):
\begin{widetext}
\begin{eqnarray}
H_{x-open}=\sum c_{x+1}^{\dag}\frac{\tau_z\sigma_0+i\tau_x\sigma_x}{2}c_x+c_{x-1}^{\dag}\frac{\tau_z\sigma_0-i\tau_x\sigma_x}{2}c_x+c_x^{\dag}\large[(2-m)\tau_z\sigma_0+b\tau_0\sigma_z+i\delta\tau_x\sigma_0\large]c_x.
\end{eqnarray}
\end{widetext}
Here $c_x$ denotes the annihilation operator for the $x$-th lattice site. Suppose there are $L$ unit cells along $x$ direction. We take the trial wave function for the $(\bar{1}00)$ surface state (i.e., localized at $x=1$) as
\begin{eqnarray}
|\chi_{\bar{1}00}\rangle=\sum_x\beta_1^x|x\rangle|\phi_1\rangle
\end{eqnarray}
where $|\phi_1\rangle$ is the spinor part and $|\beta_1|<1$.  At site $x$, the Harper equation is ($\Gamma_0=\tau_z\sigma_0$)
\begin{widetext}
\begin{eqnarray}\label{harper}
\Gamma_0\large[\frac{1-\tau_y\sigma_x}{2}\beta_1^{-1}+\frac{1+\tau_y\sigma_x}{2}\beta_1+(2-m+b\tau_z\sigma_z-\delta\tau_y\sigma_0)\large]|\phi_1\rangle=0,
\end{eqnarray}
\end{widetext}

In the above equation, we have assumed the surface-state energy to be zero, which will be validated at the end of the discussion. The $b$ term and $\delta$ term in the parentheses commute with $\tau_y\sigma_x$. The eigenstates of $\tau_y\sigma_x$ with eigenvalue $+1$ are
\begin{eqnarray}
|+1\rangle=\frac{(-i,0,0,1)^T}{\sqrt{2}};~~|+2\rangle=\frac{(0,-i,1,0)^T}{\sqrt{2}}.
\end{eqnarray}
The eigenstates of $\tau_y\sigma_x$ with eigenvalue $-1$ are
\begin{eqnarray}
|-1\rangle=\frac{(i,0,0,1)^T}{\sqrt{2}};~~|-2\rangle=\frac{(0,i,1,0)^T}{\sqrt{2}}.
\end{eqnarray}
It is easy to see from Eq. (\ref{harper}) that $|+1\rangle$ and $|+2\rangle$ can be taken as the basis of the $(\bar{1}00)$ surface states. We assume the spinor part of the solution to be
\begin{eqnarray}
|\phi_1\rangle=p_1|+1\rangle+p_2|+2\rangle.
\end{eqnarray}
Combing the normalization condition $|p_1|^2+|p_2|^2=1$, we set $p_1=\cos\theta$, $p_2=\sin\theta e^{i\phi}$, the Harper equation reduces to following complex equations 
\begin{eqnarray}
(\beta_1-m+2+b)\cos\theta-\delta\sin\theta e^{i\phi}&=&0;\notag\\
-\delta\cos\theta+(\beta_1-m+2-b)\sin\theta e^{i\phi}&=&0.
\end{eqnarray}
The solutions are given by $\beta_1=-\sqrt{\delta^2+b^2}+m-2$ (note $|\beta_1|<1$ is required for the $(\bar{1}00)$ surface), $\theta=\arctan\frac{\beta_1-m+2+b}{\delta}$, and $\phi=0$. For the $(100)$ surface, we take the trial wave function as
\begin{eqnarray}
|\chi_{100}\rangle=\sum_x\beta_2^{L-x}|x\rangle|\phi_2\rangle,
\end{eqnarray}
where $|\phi_2\rangle$ denotes the spinor part. $|-1\rangle$ and $|-2\rangle$ can be taken as the basis of the $(100)$ surface states. Similar procedure yields the solution of the Harper equation. To summarize, we have the following surface states solutions (neglecting the total normalization factor)
\begin{eqnarray}
|\chi_{\bar{1}00}\rangle&\sim&\sum_x \beta^x|x\rangle\large[\cos\theta|+1\rangle+\sin\theta|+2\rangle\large];\\
|\chi_{100}\rangle&\sim&\sum_x\beta^{L-x}|x\rangle\large[\cos\theta|-1\rangle-\sin\theta|-2\rangle\large].
\end{eqnarray}

Now we are ready to work out the surface states of a finite-size system along $x$ direction. For a finite $x$-layer, the top and bottom surface states couple together. The surface modes should be the superposition of both $|\chi_{\bar{1}00}\rangle$ and $|\chi_{100}\rangle$ and simultaneously localized on both $x=1$ and $x=L$. It is easy to calculate the finite-layer coupling:
\begin{eqnarray}
&&\langle\chi_{100}|H_{x-open}|\chi_{100}\rangle=\langle|\chi_{\bar{1}00}|H_{x-open}|\chi_{\bar{1}00}\rangle=0;\notag\\
&&\langle\chi_{100}|H_{x-open}|\chi_{\bar{1}00}\rangle=\langle|\chi_{\bar{1}00}|H_{x-open}|\chi_{100}\rangle\notag\\
&&\sim\beta^L[-b-(2-m)].
\end{eqnarray}
The small off-diagonal term (scale as $\beta^L$) will pin the surface state to be the superposition of $|\chi_{100}\rangle$ and $|\chi_{\bar{1}00}\rangle$ as
\begin{eqnarray}
|\chi_{\pm}\rangle=\frac{|\chi_{100}\rangle \pm|\chi_{\bar{1}00}\rangle}{\sqrt{2}}.
\end{eqnarray}

In the following, we consider the effect of nonzero but small $k_y$, $k_z$ terms. To be concise, we only consider the spinor part and neglect the total normalization factor of $|\chi_{\bar{1}00}\rangle$ and $|\chi_{100}\rangle$. For the $k_y$ term, we have the following relations: $\langle+1|\tau_x\sigma_y|+1\rangle=-\langle+2|\tau_x\sigma_y|+2\rangle=-\langle-1|\tau_x\sigma_y|-1\rangle=\langle-2|\tau_x\sigma_y|-2\rangle=1$ and other terms are zero. Hence $\langle\chi_{\bar{1}00}|\tau_x\sigma_y|\chi_{\bar{1}00}\rangle=-\langle\chi_{100}|\tau_x\sigma_y|\chi_{100}\rangle=\cos 2\theta$, and $\langle\chi_{\bar{1}00}|\tau_x\sigma_y|\chi_{100}\rangle=\langle\chi_{100}|\tau_x\sigma_y|\chi_{\bar{1}00}\rangle=0$. In the surface-state subspace spanned by $|\chi_{\bar{1}00}\rangle$ and $|\chi_{100}\rangle$, the $k_y$ term yields an energy splitting proportional to $\pm \cos 2\theta k_y$, which would pin the surface states to be localized at one single surface.

For the $k_z$ term, $\langle+1|\tau_x\sigma_z|+2\rangle=-\langle+2|\tau_x\sigma_z|+1\rangle=-\langle-1|\tau_x\sigma_z|-2\rangle=\langle-2|\tau_x\sigma_z|-1\rangle=i$ and all other terms are zero. Unlike the $k_y$ term which is diagonal in the basis, the $k_z$ term is non-diagonal. $|\chi_{\bar{1}00}\rangle$ and $|\chi_{100}\rangle$ are not the eigenvectors of the new Hamiltonian when a nonzero $k_z$ term is included. To extract the effect of $k_z$ term, we first solve the following Harper equation without non-Hermitian $\delta$ term:
\begin{eqnarray}
&&\Gamma_0\large[\frac{1-\tau_y\sigma_x}{2}\beta_1^{-1}+\frac{1+\tau_y\sigma_x}{2}\beta_1\notag\\
&&+(1+\cos k_z-m+b\tau_z\sigma_z+i\sin k_z\tau_y\sigma_z)\large]|\phi_1\rangle=0.\notag\\
\end{eqnarray}
Following the same procedure before, we solve the zero-energy surface states of this Hermitian topological insulator. As $\{\tau_y\sigma_z,\tau_y\sigma_x\}=0$, the $\tau_y\sigma_z$ term would mix the $\pm$ subspace of $\tau_y\sigma_x$: $\tau_y\sigma_z|+1\rangle=|-2\rangle$; $\tau_y\sigma_z|+2\rangle=-|-1\rangle$; $\tau_y\sigma_z|-1\rangle=-|+2\rangle$; $\tau_y\sigma_z|-2\rangle=|+1\rangle$. We set the trivial spinor wave function for the $(\bar{1}00)$ surface to be
\begin{eqnarray}
|\phi_{H1}\rangle=\cos\theta_H|+1\rangle+\sin\theta_H e^{i\phi_H}|-2\rangle.
\end{eqnarray} 
Solving the Harper equation yields ($m'=1+\cos k_z-m$)
\begin{widetext}
\begin{eqnarray}
\beta_{H1}&=&\frac{-1+b^2-\sin^2 k_z-m'^2-\sqrt{4(b^2-m'^2)+(1-b^2+\sin^2 k_z+m'^2)^2}}{2(m'-b)},\notag\\
\theta_H&=&-\arctan\frac{\sin k_z}{\beta_{H1}^{-1}+m'+b},\notag\\
\phi_H&=&-\frac{\pi}{2}.
\end{eqnarray}
\end{widetext}
Similarly we can solve the spinor wave function for the $(100)$ surface. The solutions are list as below:
\begin{eqnarray}
&&|\phi_{H1}\rangle=\cos\theta_H|+1\rangle-i\sin\theta_H |-2\rangle;\\
&&|\phi_{H2}\rangle=\cos\theta_H|-1\rangle+i\sin\theta_H |+2\rangle.
\end{eqnarray}

Now let us consider the effect of non-Hermitian $\delta$ term on the basis $|\phi_{H1,2}\rangle$: $\langle\phi_{H1}|i\tau_x\sigma_0|\phi_{H1}\rangle=\langle\phi_{H2}|i\tau_x\sigma_0|\phi_{H2}\rangle=i\sin 2\theta_H$ and $\langle\phi_{H1}|i\tau_x\sigma_0|\phi_{H2}\rangle=\langle\phi_{H2}|i\tau_x\sigma_0|\phi_{H1}\rangle=0$. These relations mean that the non-Hermitian $\delta$ term induces an equal energy shift for both the surface states. When $k_z$ is nonzero but small, $\theta_H\propto k_z$, and the energy shift for the surface states is $\propto ik_z$. In Eq. (\ref{harper}), we have implicitly taken the surface-state energy to be zero for a finite non-Hermitian $\delta$ term. Note that when $k_z=0$, $\theta_H=0$, hence the non-Hermitian term does not change the surface-state energy for $k_z=0$. 

\section{Energy spectra and wave-packet dynamics along $z$-wire}\label{smE}
\begin{figure*}[t!]
\includegraphics[width=0.82\textwidth]{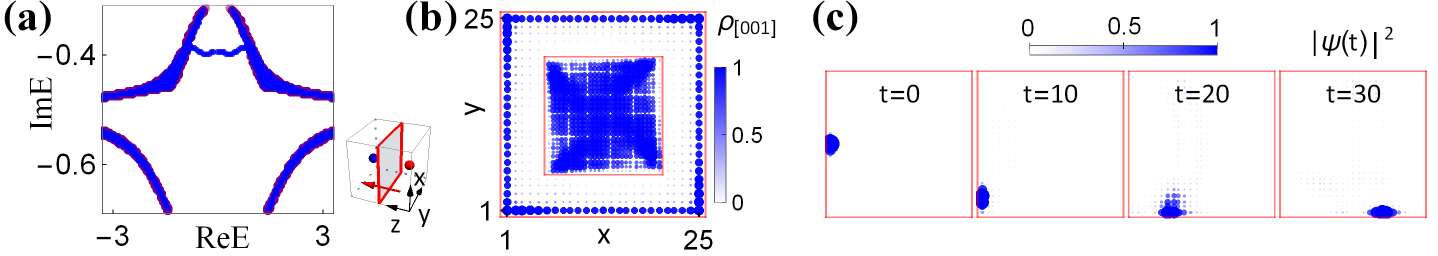}
\caption{Energy spectra and wave-packet dynamics along $z$-wire. (a) Energy spectra for a wire extending along $z$ with open (in blue) versus periodic (in purple) boundary conditions in the $x,y$ directions, showing the emergence of in-gap edge modes for open boundaries. The inset illustrates the the $z$-wire with its cross-section boundaries indicated by red lines. The lattice momentum along the wire (red arrow) is chosen as $k_z=-0.5$. (b) Total probability distribution $\rho$ of the in-gap modes, which confirms that they are localized at the edge of the cross section. The middle inset show the total probability distribution of all other modes, and there is no sign of skin effect. (c) The time evolution of a wave-packet initially localized at site $(1,13)$ with momentum $k_z=-0.5$. It undergoes chiral motion along the edges of the cross section that measures $25\times 25$ \cite{wsmanimation}. The wave-packet has width $W_1^2=2$, $W_2^2=6$. The spinor wave function is set as $|\xi_0\rangle=(1,i,0,0)^T$. Other parameters are the same as Fig. \ref{fig1}.}\label{figs4}
\end{figure*}

In the main text, we have considered the energy spectra under $[101]$-wire and the corresponding wave-packet dynamics. Here, as a comparison, we investigate energy spectra and wave-packet motion along the $z$-wire and show the anisotropic nature of non-Hermitian WSM. The spectrum of a $z$-wire with open $x,y$ boundaries is shown in Fig. \ref{figs4}(a) in color blue for a particular $k_z$. Boundary modes with energies inside the point gap are revealed by comparing to the continuum spectrum (in purple, overlaid by blue) obtained by assuming periodic boundary conditions in both the $x$- and $y$-direction. The spatial distribution $\rho$ of the in-gap modes in Fig. \ref{figs4}(b) clearly shows that they reside along the four edges. Here $\rho$ is the probability at each site $(i,j)$, $\rho(i,j)=\sum_{n}|\psi_n(i,j)|^2/\rho_{max}$, with $\rho_{max}$ the maximum value of $\rho(i,j)$ and $n$ labelling the in-gap modes shown in \ref{figs4}(a). For a given $k_z$, the 2D Hamiltonian $H_{2D}(k_x,k_y)$ describes a non-Hermitian Chern insulator. The appearance of edge modes can be predicted from the Chern number $C(k_z)$. Skin effect is absent in this geometry: the total probability distribution of the continuum (as opposed to in-gap) modes shown in the middle inset of Fig. \ref{figs4}(b) is almost a constant, in accordance with $w_{x}=w_y=0$. Here $\rho(i,j)$ is defined similarly, with $n$ summed over all continuum modes. Recently it was argued that non-Hermitian skin effect is universal: it occurs whenever the energy spectra of a 2D or 3D system take up a finite area on the complex energy plane \cite{unhse}. In point-gap WSM, the bulk spectra unavoidably occupy a finite area due to the splitting of WPs along the imaginary axis. One can check that skin modes do appear for other (e.g. diamond-shaped, not shown) geometries of the $z$-wire cross section. Such geometry-dependent skin effect is typical of many 2D and 3D non-Hermitian systems.

Fig. \ref{figs4}(c) depicts the time evolution of a wave packet initially localized at the left edge of a $z$-wire. It moves counter-clockwise along the edges [See animation in Ref. \cite{wsmanimation}]. This unambiguously demonstrates the edge modes [see Fig. \ref{figs4}(b)] are chiral. This is because the cross section of the $z$-wire, as a 2D system for fixed $k_z$,  can be regarded as a Chern insulator.

\section{Possible realization in micro-ring resonators and condensed matter materials}\label{smF}
The lattice model (see Eq. \eqref{model} in the main text) can be realized using coupled micro-ring resonators. Let us rewrite the Hamiltonian Eq. \eqref{model} in a new basis: $\tau_x\rightarrow\tau_z$, $\tau_z\rightarrow-\tau_x$; $\sigma_x\rightarrow\sigma_z$, $\sigma_z\rightarrow-\sigma_x$, which corresponds to a unitary transformation $U=e^{i\frac{\pi}{4}\tau_y\sigma_0}e^{i\frac{\pi}{4}\tau_0\sigma_y}$. In the new basis, the imaginary terms are onsite lossy terms, and the lattice model reads
\begin{widetext}
\begin{eqnarray}\label{model2}
H_\mathbf{k}=\sin k_x\tau_z\sigma_z+\sin k_y\tau_z\sigma_y-\sin k_z\tau_z\sigma_x-(\cos k_x+\cos k_y+\cos k_z-m)\tau_x\sigma_0-b\tau_0\sigma_x+i \delta \tau_z\sigma_0+i\gamma\tau_0\sigma_0.
\end{eqnarray}
\end{widetext}

\begin{figure}[h!]
\includegraphics[width=3.25in]{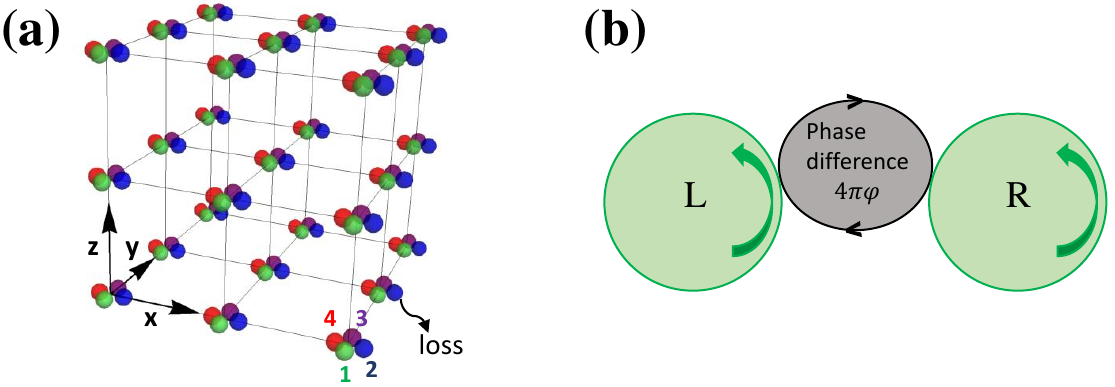}
\caption{Experimental implementation of the lattice Hamiltonian using coupled arrays of micro-ring resonators. (a) Cubic lattice formed by the micro-ring resonators. Each unit cell contains four sites, denoted by 1 (green), 2 (blue), 3 (purple), 4 (red), with loss rate $\gamma_{1,2,3,4}$, respectively. For each resonator, only the counter-clockwise (or clock-wise) propagating modes are considered. (b) Schematics of the coupling between two resonators (denoted as $L$ and $R$) through an intermediate waveguide (gray). Due to the different lengths of the upper and lower branch, a phase difference $4\pi\varphi$ is induced.}\label{figs5}
\end{figure}

We consider a 3D cubic lattice formed by ring resonators, as depicted in Fig. \ref{figs5}(a). Each unit cell consists of four ring resonators (denoted by different colors and numbered $1,2,3,4$), to mimic the $2\times 2$ orbital and spin degrees of freedom. In our notation, the $\tau_z=1$ subspace corresponds to $\{1,2\}$ sites; $\tau_z=-1$ subspace corresponds to $\{3,4\}$ sites. $\sigma_z=+1$ subspace corresponds to $\{1,3\}$ sites; $\sigma_z=-1$ subspace corresponds to $\{2,4\}$ sites. The resonators have the same resonant frequency and different loss rates, denoted as $\gamma_{1,2,3,4}$, respectively. For our case, we set $\gamma_1=\gamma_2\neq\gamma_3=\gamma_4$. The Hamiltonian Eq. \eqref{model2} contains both inter-cell and intra-cell couplings. The key ingredient implementing the couplings between two resonators is the intermediate connecting ring \cite{ringcavity1,ringcavity2} as depicted in Fig. \ref{figs5}(b). The corresponding Hamiltonian describing the couplings of the two resonators (labeled by $L$ and $R$) takes the following form:
\begin{eqnarray}\label{coupling}
-\kappa a_R^{\dag}a_L e^{i2\pi\varphi}-\kappa a_L^{\dag}a_R e^{-i2\pi\varphi},
\end{eqnarray}
where $a_{L/R}$ represents the annihilation operator of optical modes in the left/right resonator. $\kappa$ is the coupling rate and can be tuned by the overlapping between waveguide modes. $4\pi\varphi$ is the propagating phase difference inside the connecting ring, coming from the different lengths of the upper and lower branches. The phase $\varphi$ can be adjusted through, e.g., changing the length (or the refraction index) of the connecting waveguides \cite{ringcavity1,ringcavity2}.

Through the intermediate waveguide, all terms in Hamiltonian \eqref{model2} can be realized. For the inter-cell couplings, we take $\sin k_x\tau_z\sigma_z$ term as an example. Similar discussions apply to the other terms. We rewrite this term in real space:
\begin{widetext}
\begin{eqnarray}
&&\sum_{\bm r}-\frac{i}{2}\large[a_{1}^{\dag}(x+1,y,z)a_1(x,y,z)-a_{1}^{\dag}(x,y,z)a_1(x+1,y,z)\large]+\frac{i}{2}\large[a_{2}^{\dag}(x+1,y,z)a_2(x,y,z)-a_{2}^{\dag}(x,y,z)a_2(x+1,y,z)\large]\notag\\
&&+\frac{i}{2}\large[a_{3}^{\dag}(x+1,y,z)a_3(x,y,z)-a_{3}^{\dag}(x,y,z)a_3(x+1,y,z)\large]-\frac{i}{2}\large[a_{4}^{\dag}(x+1,y,z)a_4(x,y,z)-a_4^{\dag}(x,y,z)a_4(x+1,y,z)\large].\notag\\
\end{eqnarray}
\end{widetext}
Here the summation is over the unit cells $\bm r=(x,y,z)$. The subscript labels the lattice site inside each unit cell. For example, the first term represents the coupling between site-1 (green color in Fig \ref{figs5}(a)) at nearest unit cells along $x$ direction. It is easy to see from Eq. \eqref{coupling} that, this term can be reproduced by setting $\varphi=\frac{1}{4}$. Similarly, we can reproduce the other three terms by simply adjusting the phase difference of the intermediate waveguides as $\varphi=-\frac{1}{4}$, $-\frac{1}{4}$, and $\frac{1}{4}$, respectively. For the intra-cell coupling term, we take $-b\tau_0\sigma_x$ term as an example. In real space, this term is expanded as:
\begin{widetext}
\begin{eqnarray}
\sum_{\bm r}-b\large[a_{1}^{\dag}(x,y,z)a_2(x,y,z)+a_{2}^{\dag}(x,y,z)a_1(x,y,z)\large]-b\large[a_{3}^{\dag}(x,y,z)a_4(x,y,z)+a_{4}^{\dag}(x,y,z)a_3(x,y,z)\large].
\end{eqnarray}
\end{widetext}
To realize this term, we can set the phase difference of the intermediate waveguide (connecting $1,2$ or $3,4$ inside the same unit cell) as $\varphi=0$.

In practice, the 3D configuration does not require arranging the resonators on the cubic lattice. All one needs is to establish the connectivity (coordinate number) of the resonators. Also, it is worth mentioning that instead of coupling together multiple resonators to form a genuine 3D lattice, one can utilize the so-called synthetic dimension \cite{sdimen1,sdimen2,sdimen3,sdimen4,sdimen5}, e.g., the equally-spaced resonant frequency, to effectively realize the 3D lattice model on a 2D resonator array. The couplings between the multiple resonances are implemented through external modulation \cite{dmodulation} and applying the external perturbation corresponds to choosing the lattice coupling scheme and the gauge fields.

Besides micro-ring resonators, the lattice model can also be mimicked using electric circuits, where the NH Hamiltonians can be simulated by the admittance matrix. In condensed matter materials, the non-Hermitian dissipation terms can be implemented either through a tailored orbital-dependent coupling with a lossy mode or electron-phonon scattering \cite{eti}. For the case of coupling to an additional $f$-orbital, when the $f$-electron has no dispersion and sits close to the chemical potential, an effective non-Hermitian term of the form as in Eq. (\ref{model}) dominates. In a recent work on Kondo-Weyl semimetal \cite{weylfinite3} (candidate material \ce{Ce3Bi4Pd3}) which contains strongly correlated localized $f$ electrons and itinerant conduction electrons in a zincblende lattice, DMFT studies revealed that due to the breaking of inversion symmetry, the quasiparticle lifetimes at different sublattices are distinct. For the case of electron-phonon couplings, at low energies (on the scale of the point gap, measured from the energy of the WPs), the imaginary part of the electron self-energy is approximately a constant but depends on momentum and hence differs at the two WPs. Since Weyl materials typically have strong spin-orbit coupling, the anisotropy (or momentum dependence) of the lifetime is natural when there is a spin imbalance in the bath to which the electrons are coupled, such as in magnetic Weyl semimetals \cite{mweyl}.

\section{Observation of the dynamical effects}\label{smG}
As discussed in the main text, the dynamical charge pumping effect comes from the two chiral Landau levels with mismatched dissipation rates. The effective magnetic field for photons is equivalent to the complex, position-dependent coupling. For example, we can take the magnetic field $\bm{B}=B\hat{y}$ along $y$ direction and its associated gauge potential $\bm{A}=(0,0,Bx)$. Through Peierls substitution $\bm{k}\rightarrow \bm {k}-e\bm{A}$, the coupling along $z$ direction is replaced by a $x$-dependent phase. In coupled-resonator settings, the effective magnetic field can be fine-tuned as in Fig. \ref{figs5}(b) by adjusting the length (or refraction index) of the connecting waveguides or by dynamical modulating \cite{dmodulation} the refraction index through an electro-optic modulation on the ring resonator. To observe the complex chiral Landau levels, a continuous-wave laser light is injected into the resonator, with a tunable detuning $\delta\omega$. The complex band structures can be extracted from the momentum- and detuning-dependent transmission signal $s(\bm k, \delta\omega)$ from the output port \cite{sdimen5,spectralme1,spectralme2}.

Taking the advantage that the system parameters, in particular the dissipative terms, as well as their time-dependence (e.g., sudden quench of model parameters) can be easily and precisely controlled in photonic systems, it is promising to implement quantum dynamics and experimentally observe the dynamical effects induced by the non-Hermitian band topology. In contrast, in condensed matter materials, it is challenging to implement quantum quench or wave-packet motion detection. The topological features, including the surface Fermi arcs, the chiral Landau levels, and the boundary-skin modes, may be directly observed from the momentum-resolved spectrum measurement. In micro-ring resonators, the amplitude probability $c(t)=(c_{n_x,n_y,n_z}(t))$ serves as the wave function. Here $n=(n_x,n_y,n_z)$ is the index of lattice site. Its time evolution explicitly reads
\begin{eqnarray}
i\frac{dc(t)}{dt}=Hc(t).
\end{eqnarray}
In the main text, we have discussed the wave-packet dynamics for different system parameters and boundary conditions [see Fig. \ref{fig3}(c) and Fig. \ref{figs4}(c)].  As the wave-packet motion depends on the overlapping of the initial wave-packet with the eigenstates of the Hamiltonian, it can reveal the existence of surface Fermi arcs and boundary skin modes. These dynamical effects do not depend on the fine-tuning of the system parameter to some specific energies. For the dynamical charge pumping effect, we can prepare a sequence of initial wave packets (with each one localized mainly at one lattice site to mimic the trivial ground state) and measure the time-dependent amplitude distributions.


\begin{thebibliography}{99}
\bibitem{fnote1} Similar effects arise when the magnetic field is along the $x$ or $z$ direction.
\bibitem{fnote2} In fact, the form of the axion field $\theta=2(\mathbf{b\cdot r}-\delta it)$ implies that a promising direction is to construct an Euclidean action by a Wick rotation to imaginary time $\tau=it$.
\end{thebibliography}
\end{document}